\documentclass[
 aip,
 jap,
amsmath, 
reprint,
]{revtex4-2}
\usepackage{graphicx}
\usepackage{dcolumn}
\usepackage{bm}

\usepackage[utf8]{inputenc}
\usepackage[T1]{fontenc}
\usepackage{mathptmx}
\usepackage{etoolbox}
\usepackage{siunitx}
\usepackage{amssymb}
\usepackage{color,soul}
\usepackage{svg}
\usepackage{natbib}
\usepackage[final,nopatch=footnote]{microtype}
\usepackage{oubraces}
\usepackage{multirow}
\usepackage{wasysym}
\usepackage{xcolor}
\usepackage{oplotsymbl}
\usepackage{fdsymbol}
\usepackage{microtype}
\usepackage{pifont}

\makeatletter
\def\@email#1#2{%
 \endgroup
 \patchcmd{\titleblock@produce}
  {\frontmatter@RRAPformat}
  {\frontmatter@RRAPformat{\produce@RRAP{*#1\href{mailto:#2}{#2}}}\frontmatter@RRAPformat}
  {}{}
}%
\makeatother
\begin{document}

\preprint{AIP/123-QED}

\title{Low Resistance Non-Alloyed Ohmic Contacts to High Al Composition n-type AlGaN}
\author{Joseph E. Dill*}
\email[Authors to whom correspondence should be addressed: ]{jed296@cornell.edu, grace.xing@cornell.edu}
 \affiliation{School of Applied \& Engineering Physics, Cornell University, Ithaca, NY 14853, USA}
\author{Xianzhi Wei}
\affiliation{Department of Materials Science and Engineering, Cornell University, Ithaca, NY 14853, USA}
\author{Changkai Yu}%
\affiliation{Department of Materials Science and Engineering, Cornell University, Ithaca, NY 14853, USA}
\author{Akhansha Arvind}%
\affiliation{School of Electrical and Computer Engineering, Cornell University, Ithaca, NY 14853, USA}%
\author{Shivali Agrawal}%
\affiliation{Department of Chemical and Biomolecular Engineering, Cornell University, Ithaca, NY 14853, USA}%
\author{Debaditya Bhattacharya}
\affiliation{School of Electrical and Computer Engineering, Cornell University, Ithaca, NY 14853, USA}
\author{Keisuke Shinohara}%
\affiliation{Teledyne Scientific and Imaging, Thousand Oaks, CA, USA}%
\author{Debdeep Jena}
\author{Huili Grace Xing*}
\affiliation{School of Electrical and Computer Engineering, Cornell University, Ithaca, NY 14853, USA}
\affiliation{Department of Materials Science and Engineering, Cornell University, Ithaca, NY 14853, USA}
\affiliation{Kavli Institute at Cornell for Nanoscale Science, Cornell University, Ithaca, NY 14853, USA}
\date{\today}
\date{\today}

\begin{abstract}
 Ohmic contacts to high (>70\%) Al content n-type Al$_x$Ga$_{1-x}$N ultra-wide bandgap semiconductor layers in nitride electronic and photonic devices are typically fabricated by a lift-off process and high temperature ($>700^\circ$C) thermal alloying. These conditions often result in significant structural deformations of the fabricated structures and impose a harsh thermal budget on all other aspects of the device. Here, we report the fabrication of \textit{non-alloyed} \textit{as-deposited} ohmic contacts to 71\% n+AlGaN ($E_\text{g}\sim5.4$~eV) with a free carrier concentration of roughly $7\times 10^{19}$~cm$^{-3}$ and a resistivity of 4 - 5.5 m$\Omega$cm (among the lowest reported for Al$_{0.71}$Ga$_{0.29}$N) with linear $I-V$ characteristics and a contact resistivity of $\rho_\text{c}=(4.4\pm1.0)\times10^{-4}$~$\Omega$cm$^2$ (measured at zero voltage). Contacts with this quality are formed by two separate fabrication schemes: (i) metal-first patterning, and (ii) lift-off with an oxygen asher descum prior to metal deposition. Given the low threading dislocation density in the single-crystal AlN substrate used for epitaxy, the smooth morphology of the contacted epitaxial surface, and the non-alloyed nature of the contacts, this contact resistivity is attributed purely to thermionic field emission through the metal-semiconductor junction. Contact resistivity extraction at low current injection enables us to model these results using a thermionic field-emission model of contact resistivity, yielding a barrier height for Ti/Al$_{0.71}$Ga$_{0.29}$N of $(0.81\pm0.02)$ eV.
\end{abstract}

\maketitle

\section{Introduction}

AlGaN has attracted research interest for electronic applications due to its large breakdown field\cite{Tsao_2018, Doolittle_2023} and for ultraviolet photonic applications due to its large direct bandgap,\cite{Lang_2025} both of which require low-resistance metal-semiconductor contacts.

\begin{figure*}[t]
    \centering
    \includegraphics[width=\textwidth]{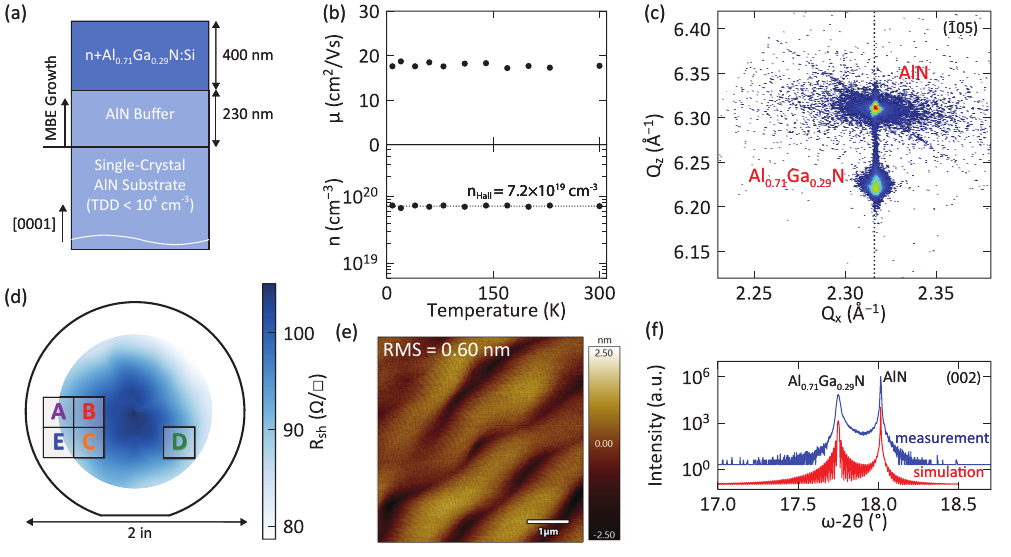}
    \caption{(a) Schematic cross-section of the n+AlGaN sample grown by molecular beam epitaxy on a 2-inch diameter metal-polar single-crystal AlN substrate\cite{Mueller_2009} (with threading dislocation density $<10^4~\text{cm}^{-2}$). (b) Mobility (top) and carrier concentration (bottom) from temperature-dependent Van der Pauw Hall measurement. (c) Reciprocal space map across the asymmetric $(\bar{1}05)$ diffractions. (d) Wafer-scale contactless sheet resistance map. Data were collected within a 0.7-inch radius of the wafer center; the edge region of the map is intentionally left blank because no data points were measured in this area. (e) $5\times5$~\textmu m$^2$ AFM scan of AlGaN surface with 0.60~nm RMS roughness. (f) Measured (blue) and simulated (red) $\omega-2\theta$ X-ray diffraction scans across the (002) diffraction.}
    \label{Fig: 1_MBE}
\end{figure*}
\begin{table*}[t]
\caption{\label{table} Hall effect characterization, fabrication specifications, and non-alloyed ohmic contact performance for the five n+Al$_{0.71}$GaN samples characterized in this study exhibiting a resistivity of 4 - 5.5 m$\Omega$cm (among the lowest reported for AlGaN with a bandgap $\sim$5.4~eV).}
\begin{ruledtabular}
\begin{tabular}{c | c c c | c c c | c c}
\multirow{2}{*}{Sample} & $\mu$ & $R_\text{sh}$ & $n$ &  {\multirow{2}{*}{Patterning}} & \multirow{2}{*}{O$_2$ descum?} &  Contact & $I-V$ & Average non-alloyed $\rho_c$\\
 & (cm$^2$/Vs) & ($\Omega/\square$) & (cm$^{-3}$) &  & & Metal & Characteristics & ($\Omega$cm$^2$)\\
\hline
A & 19 & 99  & $8.2\times10^{19}$ & Lift-off    &  No   &   Ti  &   Schottky    & --  \\
B & 19 & 118 & $6.9\times10^{19}$ & Lift-off    &  Yes  &   Ti  &   Ohmic      &   $(4.2 \pm 1.1) \times 10^{-4}$\\
C & 18 & 138 & $6.4\times10^{19}$ & Lift-off    &  No   &   V   &   Schottky    &  -- \\
D & 15 & 130 & $8.2\times10^{19}$ & Lift-off    &  Yes  &   V   &   Leaky-Schottky  &   $(13.8 \pm 4.7) \times 10^{-4}$\\
E & 19 & 107 & $7.7\times10^{19}$ & Metal-first &  No    &   Ti  &   Ohmic      &   $(7.9 \pm 2.0) \times 10^{-4}$\\
\end{tabular}
\end{ruledtabular}
\end{table*}

Ohmic contact to Si-doped n+Al$_x$Ga$_{1-x}$N has been largely reported for Al compositions up to $\sim$75\%, below which shallow doping with free electron concentrations $>10^{19}$~cm$^{-3}$ has been achieved.\cite{Mehnke_2013, Nishikawa_2023, Sarkar_2018, Bharadwaj_2019} The poor electrical conductivity obtained to date at higher Al compositions (and lack of conductive AlN substrates) is attributed to both higher dopant activation energies\cite{Gordon_2014, Trinh_2014} and compensation by acceptor-type native point defects.\cite{Nagata_2020, Almogbel_2021, Washiyama_2021}

While some AlGaN-based field-effect transistor device architectures allow for clever compositional grading to low Al composition contact surfaces,\cite{Bajaj_2016, Bajaj_2017, Xue_2019, Xue_2020, Xue_2021, Shin_2025} this luxury is not available for PN, LED, and laser diode devices, which require large epitaxial stacks without strain-induced relaxation in the active region. Hence, these device geometries typically feature a quasi-vertical configuration with a buried Al$_x$Ga$_{1-x}$N layer with Al composition $x<75\%$ to enable n-type ohmic contact formation.\cite{Zollner_2021, Kumabe_2024, Jamil_2025, Agrawal_2024, Ramesh_2025, Kushimoto_2021, Maeda_2022, Liu_2025, Bhattacharya_2025, Huang_2025}

Many recent reports of ohmic contacts to $x>70\%$ n+Al$_x$Ga$_{1-x}$N utilize vanadium (V) or titanium (Ti) metal stacks patterned by the lift-off process and annealed at temperatures exceeding $750^\circ$C.\cite{France_2007, Baca_2016, Nagata_2017, Klein_2017, Armstrong_2018, Bharadwaj_2019, Sulmoni_2020, Xue_2021, Zollner_2021, Maeda_2022, Cho_2022, Cho_2023, Ebata_2024, Kumabe_2024, Guo_2025, Liu_2025, Jamil_2025, Huang_2025, Bhattacharya_2025}
While contact resistivities approaching $1\times10^{-6}~\Omega$cm$^2$ have been reported,\cite{France_2007, Nagata_2017, Bharadwaj_2019} these high annealing temperatures typically produce significant structural deformation of the metal features,\cite{Cho_2023, Yafune_2011, Mori_2016, Klein_2017, Sulmoni_2020, Cho_2022, Nagata_2017, Ebata_2024, Guo_2025b} while also imposing harsh thermal budgets on all other components of the device heterostructure.\cite{Huang_2025}

\begin{figure*}[t]
    \centering
    \includegraphics[width=0.8\textwidth]{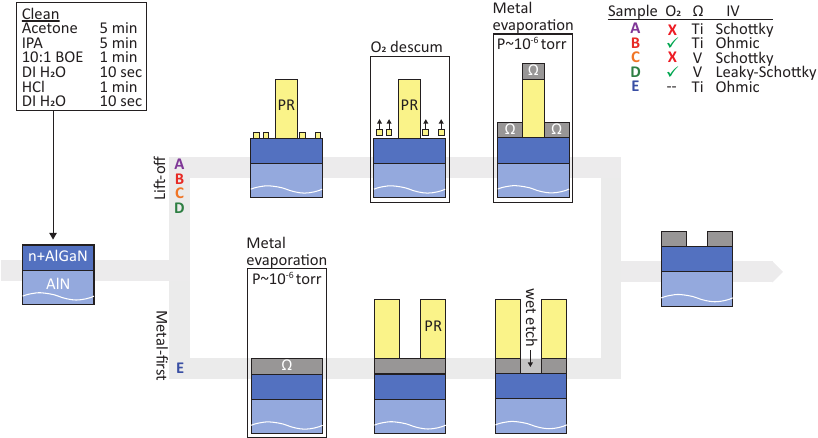}
    \caption{Device schematic and fabrication process flow for lift-off (top; samples A-D) and metal-first (bottom; sample E) ohmic contact formation. The contact metal ($\Omega$) and oxygen asher descum treatment for each sample are specified in the upper right.}
    \label{Fig: 2_fabrication}
\end{figure*}

A previous study by \citet{Smith_2023} of ohmic contacts to the wide-bandgap semiconductor $\beta$-Ga$_2$O$_3$ reported the formation of non-alloyed ohmic contacts comparable to those of alloyed contacts at similar doping densities, achieved by employing a metal-first fabrication scheme rather than a conventional lift-off method. 
A follow-up transmission electron microscopy (TEM) study\cite{Pieczulewski_2025} detected adventitious carbon at metal-semiconductor interfaces fabricated in a metal-first fashion, and a $\sim$1~nm carbon layer at interfaces patterned by a conventional lift-off process. 
Successful removal of carbon on the surface by an oxygen asher descum treatment enabled best-in-class ohmic contacts without alloying. 

Previous studies of ohmic contacts to n-AlGaN have also noted the utility of a UV/ozone descum for forming low-resistance contacts.\cite{Ebata_2024} Employing a similar strategy in this report, we demonstrate ohmic contacts to x=71\% n+Al$_x$Ga$_{1-x}$N with linear $I-V$ characteristics and $\rho_\text{c}\sim4\times10^{-4}~\Omega\text{cm}^2$ with \textit{non-alloyed, as-deposited} Ti by both a metal-first patterning scheme and a lift-off patterning scheme with proper descum removal of photoresist residue on the semiconductor surface, both of which aim to alleviate carbon contamination at the metal-semiconductor interface.

\section{Sample Preparation}

Figure \ref{Fig: 1_MBE}(a) shows the epitaxial structure of the samples used in this study. The AlGaN:Si sample was grown on a 2-inch Al-polar single crystal AlN substrate\cite{Mueller_2009} by plasma-assisted molecular beam epitaxy (MBE) in a Veeco Gen10 system equipped with standard effusion cells for Ga, Al, and Si, and a radio frequency plasma source for active nitrogen. The substrate underwent ultrasonic cleaning in solvents, followed by acid treatment, before loading into the MBE system.\cite{Lee_2020} Before the epitaxial growth, \textit{in-situ} Al-assisted surface cleaning was performed to further deoxidize the substrate surface and avoid dislocation nucleation.\cite{Cho_2020b} The epitaxial growth began with the deposition of a 230~nm~thick AlN layer under Al-rich conditions at a thermocouple temperature of $970^\circ$C. The excess Al was then desorbed \textit{in-situ}, after which the AlGaN:Si layer was grown at a temperature of $780^\circ$C under Ga-rich conditions. A Si cell temperature of $1340^\circ$C was used for a targeted Si doping density of $9\times10^{19}$~cm$^{-3}$. More details of the MBE growth of AlGaN:Si can be found in our previous report.\cite{Lee_2021}

Following the epitaxial growth, X-ray diffraction (XRD) measurements were carried out to determine the Al mole fraction and the strain state of the AlGaN layer. As shown in Fig.~\ref{Fig: 1_MBE}(c), X-ray reciprocal space mapping (RSM) of the ($\bar{1}05)$ diffraction spectra confirmed that the AlGaN epilayer was coherently strained to the underlying AlN substrate. The Al mole fraction of the AlGaN layer was determined to be $71\%$ from a symmetric XRD $\omega-2\theta$ scan along the (002) diffraction [see~Fig.~\ref{Fig: 1_MBE}(f)]. Fig.~\ref{Fig: 1_MBE}(e) shows the surface morphology of the sample measured by atomic force microscopy (AFM), exhibiting a root-mean-square (RMS) roughness of 0.60~nm over a $5\times5$~\textmu m$^2$ area. The presence of clear atomic steps indicates a step-flow growth mode, while the finger-like features are attributed to kinetically-driven instabilities induced by a pronounced Ehrlich–Schwöbel barrier.\cite{Kaufmann_2016, Lee_2021}

To evaluate the electrical conductivity of the AlGaN layer, contactless sheet resistance mapping of the two-inch wafer was performed [see~Fig.~\ref{Fig: 1_MBE}(d)]. We measured an average sheet resistance of 93~$\Omega/\square$ across the wafer, with a standard deviation of 6~$\Omega/\square$. A gradient was observed from $\sim$100~$\Omega/\square$ at the wafer's center to $\sim$80~$\Omega/\square$ near its edges, likely due to flux and temperature nonuniformity across the wafer during the MBE growth.

The two-inch wafer was diced into $8\times8$~mm$^2$ pieces for ohmic contact fabrication. Five samples (labeled A-E) were selected for this study [see~Fig.~\ref{Fig: 1_MBE}(d)]. Temperature-dependent Hall effect measurements of a diced piece from 8~K to 300~K are shown in Fig.~\ref{Fig: 1_MBE}(b). The measured carrier concentration ($7.2\times10^{19}~\text{cm}^{-3}$) remains constant with temperature, indicating degenerate Si doping above the Mott transition, which is well-suited for forming ohmic contacts. The electron mobility also remains constant with temperature at 17~cm$^2$/Vs. These transport characteristics are consistent with our previous reports of degenerately-doped n+AlGaN grown by MBE on AlN-on-sapphire templates,\cite{Bharadwaj_2019} though exhibiting $\sim2$ times higher mobility and $\sim5$ times higher carrier concentration. 

Room temperature Hall effect measurements of the five samples before fabrication [summarized in Table~\ref{table}] showed free electron concentrations ranging from $6.4$-$8.2\times10^{19}$~cm$^{-3}$ and mobilities from 15-19~cm$^2$/Vs. These n-Al$_{0.71}$GaN samples exhibit a resistivity of 4 - 5.5 m$\Omega$cm, which is among the lowest reported for this Al composition AlGaN. It is also worth noting that at the AlGaN/AlN interface, polarization discontinuity could introduce up to $1.5\times10^{13}$~cm$^{-2}$ mobile holes. However, this would be effectively masked by the $\sim2\times10^{15}$~cm$^{-2}$ electron sheet charge in these degenerately-doped AlGaN:Si layers, so it can safely be ignored in the resistivity analysis.

Circular transfer length method (C-TLM) test structures were fabricated on the diced samples, with systematic variations in the contact metal, semiconductor surface preparation, and metal patterning procedures. The fabrication procedure is outlined in Fig.~\ref{Fig: 2_fabrication}, and summarized in Table~\ref{table}. The samples were first cleaned by sonication in acetone and IPA for 5~minutes each, then soaked in 10:1 BOE solution and HCl for 1~minute each, separated by 10-second rinses in DI water. They were then blow-dried with N$_2$ for 20~seconds.

Samples A-D were fabricated using a lift-off procedure in which the n+AlGaN surface was coated with a negative photoresist, then patterned by optical lithography to open holes in the resist in which to deposit the metal stack. Samples B and D underwent a 2-minute oxygen asher descum before metalization, while samples A and C did not. The contact metal stacks, consisting of \textbf{$\Omega$}/Al/Ti (20~nm / 120~nm / 80~nm), were deposited by e-beam evaporation at a pressure of $\sim2\times10^{-6}$ torr, \textbf{$\Omega$} being the contact metal (Ti for samples A, B; V for samples C and D). 

Sample E was patterned using a metal-first scheme. After solvent and acid cleaning, a Ti/Au (20~nm / 200~nm) metal stack was deposited on the bare n+AlGaN surface. Photoresist was patterned on top of the metal to reveal the channel regions between the contacts and then hard-baked for 5~minutes at $115^\circ$C to form an etch-resistant hard mask. The Au layer was removed by submersion in Au etchant TFA, and the Ti layer was removed by submersion in a 20:1 diluted BOE solution.\cite{Smith_2024}

\begin{figure*}[t]
    \centering
    \includegraphics[width=0.95\textwidth]{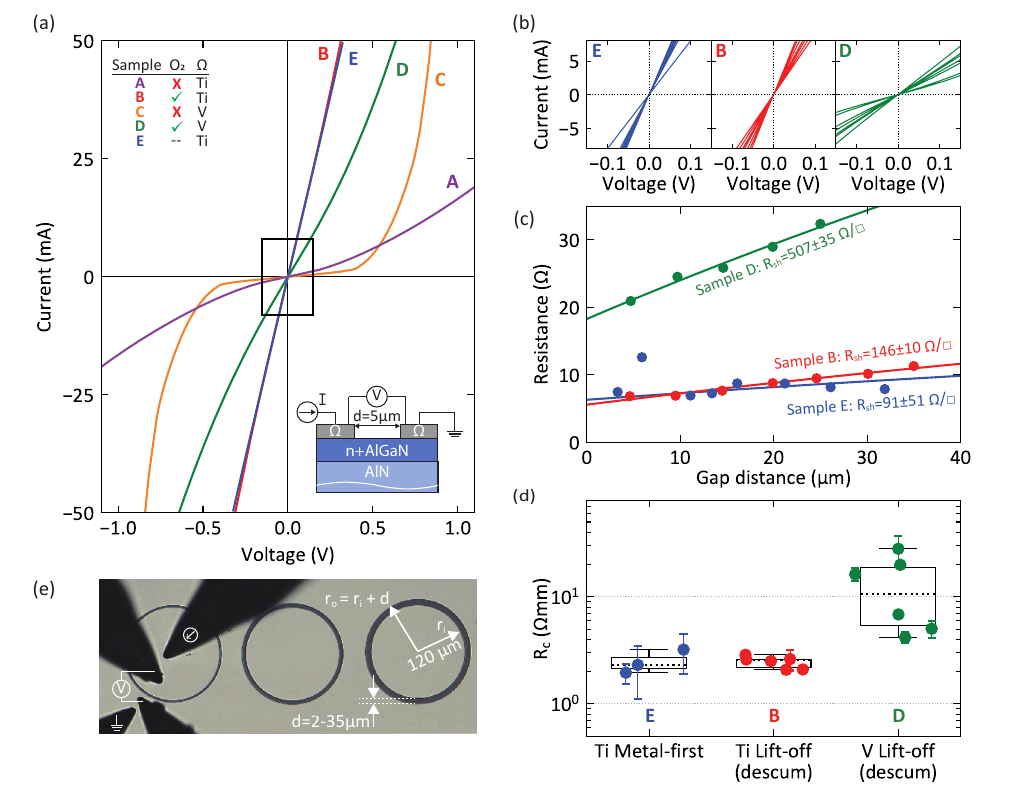}
    \caption{(a) Current-voltage ($I-V$) characteristics of circular transfer length method (C-TLM) test structures from (a) samples A-E with a gap distance of $\sim$5~\textmu m. (upper-left inset) Summary of the surface treatment and contact metal for samples A-E. (lower-right inset) Circuit configuration for the current-sourced $I-V$ measurements. (b) Samples E, B, and D in the linear $I-V$ regime with gap distances from 2-35~\textmu m. (c) TLM plot of samples E, B, and D based on linear I-V shown in (b).
    (d) Box-and-whisker plot of the contact resistance $R_\text{c}$ extracted from 0 to 5 mA from all test structures on samples E, B, and D. 
    Current-density-dependent $R_\text{c}$ analysis\cite{Piotrzkowski_2011, Hu_2015, Dill_2025, Huang_2025, Bhattacharya_2025} of the non-linear $I-V$ curves measured on samples A and C could not be performed as this analysis is \textit{only applicable to linear TLM} test structures (but not C-TLM since a C-TLM contact pair does not share the same current density). The corresponding specific contact resistivity values, averaged across all test structures on each sample, are listed in Table \ref{table} (in units of $\Omega$cm$^2$). 
    (e) (left) Microscope image of fabricated C-TLM pads probed in a four-point configuration and diagram indicating the dimensions of the inside and outside radii ($r_\text{i}$ and $r_\text{o}$, respectively) of the C-TLM test structures.}
    \label{Fig: 3_CTLM}
\end{figure*}

\section{C-TLM Measurement and Analysis}

Electrical characterization of the C-TLM test structures was performed at room temperature to assess the quality of the ohmic contact at the non-alloyed metal/n+AlGaN interfaces for the various surface preparation and metalization schemes. These measurements are summarized in Fig.~\ref{Fig: 3_CTLM} and in Table~\ref{table}. The C-TLM structures had an inner radius of $r_\text{i}=120$~\textmu m and outer radius $r_\text{o}=r_\text{i}+d$ where the gap distance $d$ ranged from 2 to 35~\textmu m [see~Fig.~\ref{Fig: 3_CTLM}(e)]. All lithographic dimensions were confirmed by optical microscope measurements under 700X magnification. DC $I-V$ measurements were performed in a four-probe configuration, sourcing a current $I$ to the inside circular pad and measuring the voltage $V$ across the semiconductor gap [see inset to Fig.~\ref{Fig: 3_CTLM}(a)].

Fig.~\ref{Fig: 3_CTLM}(a) shows the $I-V$ characteristics of select C-TLM structures with a 5~\textmu m gap from all five fabricated samples. Consistent with the previously-mentioned studies of non-alloyed ohmic contacts to $\beta$-Ga$_2$O$_3$\cite{Smith_2023, Smith_2024, Pieczulewski_2025}, the fabrication processes that mitigate the presence of carbon in between the metal and semiconductor surfaces yield the highest-performing ohmic contacts. Samples A and C, which were patterned by lift-off and did not undergo an oxygen asher descum, have highly non-linear $I-V$ characteristics. By contrast, sample E, patterned with Ti as the contact metal using metal-first processing, never acquiring any photoresist between the metal and semiconductor surface, exhibits linear $I-V$ characteristics. Sample B, which underwent a lift-off procedure but had the photoresist residue removed by the oxygen asher descum process, exhibits equivalent $I-V$ to the metal-first sample. Sample D, which has V as its contact metal, passes less current than the equivalent Ti-based sample, but significantly more current than the V-based sample that did not undergo asher descum (sample C), particularly at low current injection.

The resistance of a C-TLM test structure is given by \cite{Cohen_2014}
\begin{equation}\label{Eq: R_CTLM}
    R_\text{CTLM} = \frac{R_\text{sh}}{2\pi} \bigg[\log \Big(\frac{r_\text{o}}{r_\text{i}}\Big) + \frac{L_\text{t}}{r_\text{i}}\frac{\text{I}_0(r_\text{i}/L_\text{t})}{\text{I}_1(r_\text{i}/L_\text{t})} +\frac{L_\text{t}}{r_\text{o}}\frac{\text{K}_0(r_\text{i}/L_\text{t})}{\text{K}_1(r_\text{i}/L_\text{t})}\bigg]
\end{equation}
where $r_\text{i}$ and $r_\text{o}=r_\text{i}+d$ are the inside and outside radii, respectively, $d$ being the gap distance. $\text{I}_n(x)$ and $\text{K}_n(x)$ are $n^\text{th}$-order modified Bessel functions of the first and second kind. The transfer length $L_\text{t}=\sqrt{\rho_\text{c}/R_\text{sh}}$ corresponds to the distance inside the edge of the contacts at which the mobile electron current terminates, finding the collective path of least resistance through the semiconductor material under the contact -- assumed to have sheet resistance $R_\text{sh}$ -- and across the metal-semiconductor junction, having specific contact resistivity $\rho_\text{c}$. 

Specific contact resistivity is defined by a differential form of Ohm's law,
\begin{equation}\label{Eq: rho_c}
  \rho_\text{c} = \Big(\frac{\partial J}{\partial V} \Big)^{-1}_{V=0}
\end{equation}
where $J$ is the current density and $V$ is the voltage across the metal-semiconductor junction. This quantity, and by extension, $L_\text{t}$ and $R_\text{CTLM}$ in Eq.~\ref{Eq: R_CTLM}, are strictly defined as a derivative \textit{at zero voltage}. Thus, in C-TLM measurements, $\rho_\text{c}$ is only rightly extracted within a current-injection range in which the contacts' $I-V$ characteristics are linear through the origin. 

Analysis models have been proposed that generalize $\rho_\text{c}$ extraction for linear TLM measurements in the case of non-linear $I-V$ characteristics,\cite{Piotrzkowski_2011} or analyze non-linear C-TLM $I-V$ characteristics by directly fitting the $I-V$ curves with an equivalent-circuit model.\cite{Patel_2007} Our own earlier studies have utilized \textit{current density}-dependent L-TLM analysis to quantify the degree of the contact nonlinearity and determine its impact on device performance.\cite{Hu_2015, Dill_2025, Huang_2025, Bhattacharya_2025} The assumption behind extracting contact resistance at a given current density instead of a given voltage is that the leaky-Schottky contacts in a TLM set are spatially uniform and exhibit the same voltage drop for a given current density flowing through a pair of TLM contacts. Consequently, this assumption necessitates the use of linear TLM test structures, as the current density changes between the inner and outer contacts in C-TLM structures. 

Therefore, for the present study, we limit our contact resistance analysis to a current range of $\pm5$~mA (through C-TLM structures with an inside radius of 120~\textmu m), in which the measured $I-V$ characteristics of C-TLM structures on samples B, D, and E are linear.

We note here that C-TLM analysis by Eq.~\ref{Eq: R_CTLM} can also be obfuscated in the case of annealed contacts, where metal diffusion into the semiconductor under the contacts likely changes $R_\text{sh}$ under the contacts compared to the material between the contacts. However, because the contacts analyzed in this study are not annealed, the assumption of equivalent $R_\text{sh}$ beneath the contacts and between the contact pads is well-justified.

The resistances of the $I-V$ curves measured on samples E (Ti metal-first), B (Ti lift-off with oxygen descum), and D (V lift-off with oxygen descum) were determined by linear fits between $\pm5$~mA [see~Fig.~\ref{Fig: 3_CTLM}(b)]. The resulting resistances were fit to Eq.~\ref{Eq: R_CTLM} as a function of gap distance by an orthogonal distance regression with $L_\text{t}$ and $R_\text{sh}$ as free fitting parameters, weighted by uncertainties in both the measured resistances and the measured test structure dimensions [see~Fig.~\ref{Fig: 3_CTLM}(c)]. 

Fig.~\ref{Fig: 3_CTLM}(d) shows the spread in measurements of contact resistance, $R_\text{c}=R_\text{sh}L_\text{t}$ (in units of $\Omega$mm), for each sample. This quantity is related to the \mbox{y-intercept} of an $R_\text{CTLM}$ vs $d$ plot [i.e.,~Fig.~\ref{Fig: 3_CTLM}(c)] by $\frac{1}{2}R_\text{CTLM}(d=0)=R_\text{c}/2\pi r_\text{i}$. Samples E and B, which have Ti as their contact metal, show similar values of $R_\text{c}\sim2$~$\Omega$mm. The metal-first sample showed larger \mbox{nonuniformity} within the set of measured C-TLM test structures, evidenced by non-monotonicity of resistance with gap distance in Fig.~\ref{Fig: 3_CTLM}(c) and resulting in a larger spread of $R_\text{c}$ values with larger error bars in Fig.~\ref{Fig: 3_CTLM}(d). By contrast, the de-scummed sample (B) shows clean monotonicity in $R$ vs $d$, and smaller spread and error bars in Fig.~\ref{Fig: 3_CTLM}(d). 

\begin{figure*}[t]
    \centering
    \includegraphics[width=0.8\textwidth]{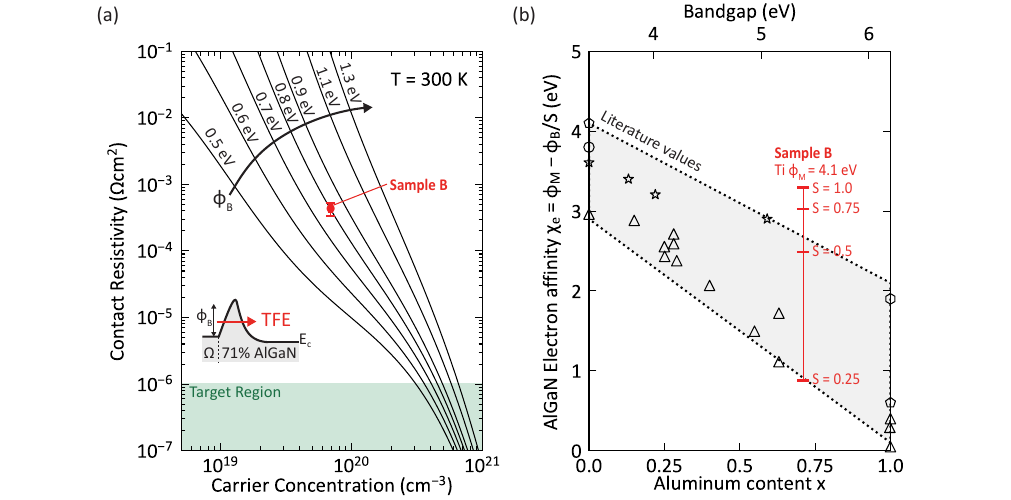}
    \caption{(a) Specific contact resistivity $\rho_\text{c}$ as a function of doping density $N_\text{d}$, simulated for x=71\% Al$_x$Ga$_{1-x}$N using the thermionic field emission model (see inset) defined in Eq.~\ref{Eq: rhoc_TFE}. The average contact resistivity [see~Fig.~\ref{Fig: 3_CTLM}(c,d)] and Hall density (see Table~\ref{table}) measured for sample B (Ti lift-off with O$_2$ descum) are plotted, suggesting a barrier height for Ti/n+Al$_{0.71}$Ga$_{0.29}$N of $\phi_\text{B}=(0.81\pm0.02$)~eV, assuming an effective mass of 0.34~$m_0$, a dielectric constant of 8.6~$\varepsilon_0$, and a bandgap of 5.4~eV for Al$_{0.71}$Ga$_{0.29}$N (see Table~\ref{Table: AlGaN Material Parameters}). (b) The inferred electron affinity $\chi_\text{e} = \phi_\text{M}-\phi_B/S$ (see Eq.~\ref{Eq: S Schottky Mott}) from thermionic field emission modeling of sample B for $S$ values ranging from 1.0 to 0.25 (red), benchmarked against measurements of Al$_x$Ga$_{1-x}$N electron affinity from literature (black), obtained from Refs. \citenum{Wu_1999}($\varhexagon$), \citenum{Levinshtein_2001} ($\pentago$), \citenum{Grabowski_2001}($\vartriangle$), \citenum{Kozawa_2000} ($\medwhitestar$), and \citenum{Lin_2012} ($\circ$).}
    \label{Fig: 4_TFE}
\end{figure*}
\begin{table}[b]
    \centering
    \begin{tabular}{| l | c | c | c | c |}
    \hline
         Parameter              &   GaN     &   AlN     &   Al$_{0.71}$Ga$_{0.29}$  \\ \hline
         Effective mass ($m_e)$ &   0.20    &   0.40    &   0.34                    \\ \hline
         Dielectric Constant    &   8.9     &   8.5     &   8.6                     \\ \hline
         Bandgap (eV)           &   3.4     &   6.2     &   5.4                     \\ \hline
    \end{tabular}
    \caption{Nitride material parameters used in thermionic field emission model, obtained from Ref.~\citenum{Levinshtein_2001}. The AlGaN values are inferred by linear interpolation between those of GaN and AlN with two significant digits.}
    \label{Table: AlGaN Material Parameters}
\end{table}

These results imply that the oxygen asher descum effectively creates a uniform AlGaN contact surface. Additionally, we conclude that the electrical properties of the de-scummed sample approximate an ideal Ti/AlGaN interface, given the agreement in measured $R_\text{c}$ between sample C and the metal-first sample (E). While the high nonuniformity observed in this sample prohibits definitive, low-uncertainty C-TLM analysis, the linear $I-V$ characteristics [Fig. \ref{Fig: 3_CTLM}(a)] and low overall resistance of the C-TLM test structures at low current injection [see Fig.~\ref{Fig: 3_CTLM}(c)] in this sample indicate a small contact resistance. This nonuniformity likely could have been mitigated by performing an oxygen asher descum prior to the metal-first fabrication, as recently demonstrated by Pieczulewski and Smith \textit{et al.} on $\beta$-Ga$_2$O$_3$ \cite{Pieczulewski_2025}.

The V-based sample that underwent oxygen asher descum (D) showed slightly non-ohmic $I-V$ characteristics and exhibited significantly larger variation across the sample [see Fig.~\ref{Fig: 3_CTLM}(d)]. Previous reports of \textit{alloyed} contacts to high Al content n+AlGaN have suggested superior performance by V-based metal stacks over and against Ti-based stacks.\cite{France_2007, Haidet_2017} The superior electrical performance of the Ti-based metal stacks in this study of non-alloyed contacts may arise due to the presence of a residual surface oxide on the AlGaN surface. The Ti is more effective than V at gettering this oxide, leading to a lower semiconductor bandgap at the metal-semiconductor interface. The C-TLM analysis of sample D in its linear regime yielded $R_\text{sh}$ values on the order of 500~$\Omega/\square$, significantly larger than that measured by Hall (see Table~\ref{table}) and by contactless sheet resistance mapping [see~Fig.~\ref{Fig: 1_MBE}(d)]. The origin of the large $R_\text{sh}$ in this C-TLM analysis is not well understood.

\begin{figure}[t]
    \centering
    \includegraphics[width=0.99\columnwidth]{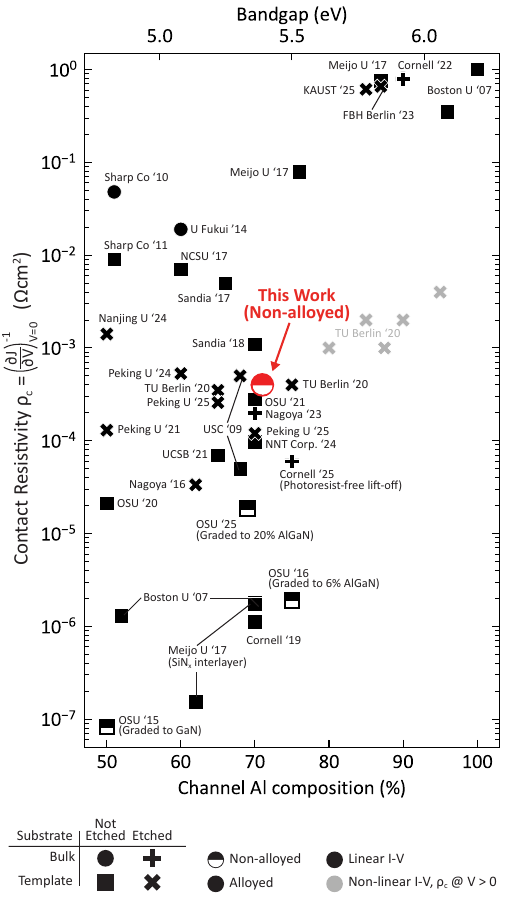}
    \caption{Benchmark plot of contact resistivity $\rho_\text{c} = \big(\frac{\partial J}{\partial V}\big)^{-1}_{V=0}$, extracted at zero voltage, versus Al content from 50\% to 100\%. Data are taken from Refs. \citenum{France_2007, Srivastava_2009, Tokuda_2010, Yafune_2011, Yafune_2014, Park_2015, Mori_2016, Bajaj_2016, Nagata_2017, Douglas_2017, Haidet_2017, Armstrong_2018, Bharadwaj_2019, Sulmoni_2020, Cho_2020, Xue_2020, Zollner_2021, Xue_2021, Zhang_2021, Maeda_2022, Cho_2023, Kumabe_2024, Zhou_2024, Guo_2024, Ebata_2024, Liu_2025, Guo_2025, Bhattacharya_2025, Guo_2025b, Shin_2025}. TLM analyses with linear $I-V$ characteristics are plotted in full opacity. When possible, for publications that show TLM analysis with non-linear $I-V$ characteristics but report a $\rho_\text{c}$ value extracted at a non-zero current injection, the provided TLM data are scrubbed and re-fit at $V=0$. Where insufficient data are provided in the manuscript to perform such re-fitting, the reported contact resistivity is plotted, but with reduced opacity. $ \circletfill/\blacksquare$ symbols mark contacts to unetched n-AlGaN surfaces, while \ding{58}/\ding{54} symbols mark those on etched surfaces for bulk/template substrates. Alloyed/non-alloyed contacts are indicated with full/top-filled markers.}
    \label{Fig: 5_Benchmark}
\end{figure}

\section{Thermionic Field Emission Modeling}

Due to the smooth morphology of the un-etched epi-grown AlGaN surface [see~Fig.~\ref{Fig: 1_MBE}(e)], low threading dislocation density of the single-crystal AlN substrate,\cite{Mueller_2009} and the absence of annealing-induced morphological defects, and high uniformity of contact performance across the sample, the linear ohmic contact resistances in the Ti-based sample B is expected to arise strictly from quantum tunneling and thermal processes across the metal-semiconductor junction, unassisted by defects or dislocations of any kind, as seen, for example, on AlGaN grown on sapphire substrates.\cite{Haidet_2015, Nagata_2017, Bharadwaj_2019} Thus, the contact resistivity of the as-deposited contacts at low current injection can be modeled using a thermionic field emission model.\cite{Padovani_1966, Murphy_1956, Crowell_1969} The modified expression derived by \citet{Smith_2024},
\begin{widetext}
\begin{equation}\label{Eq: rhoc_TFE}
  \rho_{c,\text{TFE}}(\phi_\text{B}, N_\text{d}) =
\frac{k_\text{B} \sqrt{E_{00}} \,\cosh\!\left(\frac{E_{00}}{k_\text{B} T}\right) 
\,\coth\!\left(\frac{E_{00}}{k_\text{B} T}\right)}
{A^* T q \sqrt{\pi\!\left(q\phi_\text{B} + \Delta E_{\text{BM}} - \Delta E_{\text{BGR}}\right)}}
\exp\!\left[
\frac{q\phi_\text{B} + \Delta E_{\text{BM}} - \Delta E_{\text{BGR}}}
{E_{00} \coth\!\left(\frac{E_{00}}{k_\text{B} T}\right)}
- \frac{\Delta E_{\text{BM}} - \Delta E_{\text{BGR}}}{k_\text{B} T}
\right],
\end{equation}
\end{widetext}
accounts for shifts in the optical bandgap at high doping levels due to the Burstein Moss effect ($\Delta E_\text{BM}$), which raises the bandgap due to band filling above the Gamma point in degenerately-doped semiconductors, and Bandgap renormalization ($\Delta E_\text{BGR}$), which arises from many-particle electron-electron and electron-impurity interactions, lowering the optical bandgap. These effects have previously been shown\cite{Bharadwaj_2019} to shift the bandgap of degenerately-doped Al$_{0.7}$Ga$_{0.3}$:Si by $\sim0.2$~eV.

This model is a function only of the metal-semiconductor barrier height $\phi_\text{B}$, doping density $N_\text{d}$, and material properties of the semiconductor. $k_\text{B}$ is Boltzmann's constant, $T$ is temperature, $q$ is the electron charge, and
\mbox{$A^*=4\pi qm^*k_\text{B}^2/h^3$} is Richardson's constant, where $h$ is Planck's constant. \mbox{$E_{00}=(qh/4\pi)\sqrt{n/m^*\varepsilon_\text{s}}$} is a characteristic field, with $\varepsilon_\text{s}$ the dielectric permittivity of the semiconductor. $m^*$ is the Fermi-level effective mass, corrected to account for conduction band non-parabolicity at high doping levels. Complete definitions of all variables in this model are provided in the supplementary information of Ref. \citenum{Smith_2024}.

To model Al$_x$Ga$_{1-x}$N, the uncorrected effective mass, dielectric permittivity, and bandgap were all inferred by linear interpolation between literature values for GaN and AlN, summarized in Table~\ref{Table: AlGaN Material Parameters}. 

In Fig.~\ref{Fig: 4_TFE}(a), Eq.~\ref{Eq: rhoc_TFE} is plotted vs $N_\text{d}$ for $x=71\%$ Al composition Al$_x$Ga$_{1-x}$N with contours at values of $\phi_\text{B}$ ranging from 0.4~eV to 1.1~eV.
The average measured contact resistivity $\rho_\text{c}=L_\text{t}^2R_\text{sh}$ from C-TLM analysis of sample B (Ti contact metal by lift-off with an oxygen asher descum) is benchmarked against the TFE model with $N_\text{d}$ inferred from Hall effect measurement of the sample before fabrication (see Table~\ref{table}).

Solving Eq.~\ref{Eq: rho_c} numerically for $\phi_\text{B}$ using the measured value of $\rho_\text{c}$ and extrapolating uncertainties from (i) the standard deviation of $\rho_\text{c}$ from C-TLM measurements across sample B and (ii) $N_\text{d}$ from nonuniformity between the measured samples (see Table~\ref{table}), we infer for Ti/Al$_{0.71}$Ga$_{0.29}$N a barrier height of $\phi_\text{B}=(0.81\pm0.02)$~eV. 

We note here that the barrier height extracted in this manner depends on the assumed electron effective mass via $A^*$ and $E_{00}$ (see Eq.~\ref{Eq: rhoc_TFE}). For effective mass values between 0.25~$m_\text{e}$ and 0.40~$m_\text{e}$, the extracted barrier height varies from 1.0~eV to 0.7~eV. However, for the present analysis, we assume an effective mass value of 0.34~$m_\text{e}$ (per Ref.~\citenum{Levinshtein_2001}; see Table~\ref{Table: AlGaN Material Parameters}).

The typical Schottky-Mott rule can be modified\cite{Tung_2014, Smith_2024} to account for surface Fermi level pinning as follows,
\begin{equation}\label{Eq: S Schottky Mott}
    \phi_\text{B} = S(\phi_\text{M} - \chi_\text{e})
\end{equation}
where $\phi_\text{M}$ is the work function of the contact metal and $\chi_\text{e}$ is the electron affinity of the semiconductor. \mbox{$S\in[0,1]$} is a phenomenological "index of surface behavior". An $S$ value of 1 recovers the typical Schottky-Mott relation, while an $S$ value of 0 corresponds to total surface Fermi level pinning at the conduction band.

Taking 4.1~eV to be the work function of Ti, we infer an electron affinity of $(3.28\pm0.3)$~eV for Al$_{0.71}$Ga$_{0.29}$N in the ideal Schottky-Mott case ($S=1$). This value is plotted in Fig.~\ref{Fig: 4_TFE}(b), alongside various measurements of Al$_x$Ga$_x$N electron affinity as a function of Al-content $x$ from literature.\cite{Wu_1999, Levinshtein_2001, Grabowski_2001, Kozawa_2000, Lin_2012} Our inferred electron affinity for $S=1$ lies above the range of measured values by $\sim0.5$~eV. This deviation can potentially be attributed to surface Fermi level pinning ($S<1$) at the AlGaN surface or to the unknown work function of partially-oxidized Ti in contact with AlGaN. As seen in Fig.~\ref{Fig: 4_TFE}, assuming Fermi-level pinning between $S=0.25$ and $S=0.5$, the measured barrier height corresponds to electron affinities between 2.5~eV and 0.8~eV, respectively, for Al$_{0.71}$Ga$_{0.29}$, which is within the range of measurements reported in the literature.

\section{Benchmarking}

In Fig.~\ref{Fig: 5_Benchmark}, the average contact resistivity measured on sample B (Ti lift-off with oxygen asher descum) is benchmarked against other published measurements of contact resistivity to n-type AlGaN versus the Al composition of the channel between the contacts from 50\% to 100\%. Contacts to unetched n-AlGaN surfaces are indicated with $ \circletfill/\blacksquare$ symbols, while contacts to etched surfaces are indicated with \ding{58}/\ding{54} symbols for bulk/template substrates, respectively. Alloyed contacts are marked with full-filled shapes, while non-alloyed contacts are marked with top-filled shapes.

Many TLM studies of contacting high Al content AlGaN exhibit non-linear $I-V$ characteristics and report a value of $\rho_\text{c}$ extracted at non-zero voltage bias. Keeping with the definition of contact resistivity (see Eq.~\ref{Eq: rho_c}) and to make a unilaterally fair comparison between reports, non-linear $I-V$ data were scrubbed and re-fit to obtain $\rho_\text{c}$ at $V=0$. Reports that either (i) demonstrated linear $I-V$ characteristics, or (ii) provided sufficient data to evaluate the resistance from $I-V$ characteristics about the origin and redo the TLM fitting, are marked in Fig.~\ref{Fig: 5_Benchmark} in full opacity. For reports that exhibited non-linear $I-V$ characteristics but could not be re-fit at $V=0$ (per Eq.~\ref{Eq: rho_c}), the reported $\rho_\text{c}$ values are indicated, but with reduced opacity.

In Fig.~\ref{Fig: 5_Benchmark}, the non-alloyed contacts to unetched n-type AlGaN on a bulk substrate from this report are seen to be on par with alloyed contacts to etched AlGaN with similar Al composition on both bulk and template substrates. Two reports of non-alloyed contacts to high Al content n-type AlGaN channels (Refs.~\citenum{Park_2015} and \citenum{Bajaj_2016}), achieve contacts resistivities of $\rho_\text{c}\sim10^{-7}~\Omega\text{cm}^2$ and $\rho_\text{c}\sim10^{-6}~\Omega\text{cm}^2$ to 50\% and 75\% Al content n-type AlGaN channels, respectively, by utilizing a reverse-graded architecture such that to Al content at the metal-semiconductor interface is below 6\%. The only reports\cite{France_2007, Nagata_2017, Bharadwaj_2019} that achieve comparable contact resistivity with high Al content at the metal-semiconductor interface do so with alloyed contacts to un-etched AlGaN on template substrates. Notably, the lowest-reported contact resistivity to \text{etched} n-type AlGaN above 65\% Al composition utilizes a novel photoresist-free lift-off scheme.\cite{Bhattacharya_2025}

\section{Conclusions}

In summary, we have prepared non-alloyed Ti-based ohmic contacts to degenerately-doped ($N_\text{d}\sim7\times10^{19}$~cm$^{-3}$) Al$_{0.71}$Ga$_{0.29}$N:Si by metal-first processing and by a lift-off process with oxygen asher descum prior to metalization, both of which yield linear $I-V$ characteristics and a contact resistivities of $\sim4\times10^{-4}~\Omega\text{cm}^2$. The congruence between these results suggests that the removal of surface carbon, whether adventitious or from photoresist residue, is crucial for forming ideal, electrically transparent metal/n+AlGaN interfaces, consistent with previous findings for $\beta$-Ga$_2$O$_3$.\cite{Smith_2023, Smith_2024, Pieczulewski_2025} Thermionic field emission modeling of contact resistivity,\cite{Smith_2024} accounting for Burstein-Moss and bandgap renormalization effects in the degenerately-doped n+AlGaN,\cite{Bharadwaj_2019} implies a barrier height of ($0.81\pm0.02$)~eV. This barrier height corresponds with literature values for Al$_{x}$Ga$_{1-x}$N electron affinity under the assumption of significant surface Fermi-level pinning\cite{Tung_2014} ($0.25<S<0.5$).

The analysis presented in this work can be expanded by fabricating carbon-free metal-semiconductor contacts with a range of metal work functions to fit with Eq.~\ref{Eq: S Schottky Mott} and assess the degree of Fermi level pinning on the AlGaN surface, as done in Ref.~\citenum{Smith_2024} for $\beta$-Ga$_2$O$_3$. However, it is worth noting that that appreciable barrier heights (>20 $k_\text{B}T$) can be extracted with much higher accuracy from measurements of thermionic emission (TE) current through canonical Schottky junctions fabricated on lightly doped semiconductors rather than TFE-enabled ohmic contacts fabricated on a heavily doped semiconductor. 
Such experiments must invoke the correct carrier transport model when extracting barrier height; a universal transition electric field between TE and TFE in Schottky junctions has been derived by \citet{Li_2020} along with an example analysis on barrier height extraction.

The quality of contact resistivity achieved in this work with as-deposited contacts on an as-grown surface suggests that fabrication schemes for nitride devices with buried n+AlGaN contact layers should minimize carbon contamination and surface state generation on the contacted n+AlGaN layer prior to metal deposition. For instance, eliminating photoresist contamination of the contacted n+AlGaN surface,\cite{Bhattacharya_2025} and using an etch treatment that preserves near-as-grown surface quality,\cite{Liu_2025} can both facilitate improved ohmic contact formation.

\section{Acknowledgements}
This work was supported in part by the DARPA UWBGS program under Award No: HR001123S0051-FP-26 [sample fabrication and characterization], ULTRA, an Energy Frontier Research Center funded by the U.S. Department of Energy (DOE), Office of Science, Basic Energy Sciences (BES), under Award No. DE-SC0021230 [sample growth], and by SUPREME, one of seven centers in JUMP 2.0, a Semiconductor Research Corporation (SRC) program sponsored by DARPA [modeling]. This work was performed in part at the Cornell NanoScale Facility, a member of the National Nanotechnology Coordinated Infrastructure (NNCI), which is supported by the National Science Foundation (Grant NNCI-2025233).
The authors benefited from discussions of these results with Dr. Kathleen Smith and Sheena Huang.

\section{Author Declarations}
\subsection{Conflict of Interest}
The authors have no conflicts to disclose.

\subsection{Author Contributions}
J. E. Dill and X. Wei contributed equally to this work.\\

\textbf{Joseph E. Dill}: Conceptualization (supporting), Data Curation (supporting), Formal Analysis (equal), Methodology (equal), Software (supporting), Validation (lead), Visualization (lead), Writing/Original Draft Preparation (lead); \textbf{Xianzhi Wei}: Data Curation (lead), Formal Analysis (equal), Methodology (equal), Resources (equal), Validation (supporting), Visualization (supporting); \textbf{Changkai Yu}: Resources (equal), Writing/Original Draft Preparation (supporting), Writing/Review \& Editing (supporting); \textbf{Akhansha Arvind}: Data Curation (supporting), Formal Analysis (supporting), Software (lead), Visualization (supporting); \textbf{Shivali Agrawal}: Resources (supporting), Validation (supporting); \textbf{Debaditya Bhattacharya}: Validation (supporting), Writing/Review \& Editing (supporting); \textbf{Keisuke Shinohara}: Data Curation (supporting), Writing/Review \& Editing (supporting); \textbf{Debdeep Jena}: Funding Acquisition (equal), Supervision (supporting), Writing/Review \& Editing (supporting); \textbf{Huili Grace Xing}: Conceptualization (lead), Funding Acquisition (equal), Supervision (lead), Writing/Review \& Editing (lead).

\section{Data Availability Statement}
The data that support the findings of this study are available from the corresponding author upon reasonable request. 

\section{References}
\bibliography{references}

\end{document}